\documentstyle{article}

\date{}
\newcommand{\tr}{\,\mbox{tr}\,}

\begin{document}

\begin{center}
{\Large \bf Quantum fluctuations of chiral condensate in the
analytically regularized Nambu--Jona-Lasinio model\footnote{Talk
given at XVII International Workshop on High Energy Physics and
Quantum Field Theory (QFTHEP'2003), Samara--Saratov, 4-11 Sept.
2003 }}
\\

\vspace{4mm}

 V.E. Rochev\footnote{e-mail: rochev@mx.ihep.su} \\
 Institute for High Energy Physics \\
 Protvino, Moscow region, Russia\\
\end{center}

\begin{abstract}
 The problem of
quantum fluctuations of the chiral condensate due to
next-to-leading
 contributions is investigated.   Meson contributions to the chiral
condensate are calculated in the Nambu--Jona-Lasinio model with
the analytical (dimensional) regularization in a framework of the
mean-field expansion. The pion contribution can be significant,
while the sigma-meson contribution is small in the regime of
physical values of the model parameters. Non-pole contributions
are also estimated and found to be  small.
\end{abstract}

\section*{Introduction and Summary}

The Nambu--Jona-Lasinio model  (NJL) is a successful effective
model of quantum chromodynamics  of light hadrons in the
non-perturbative region (See \cite{kle, HatKun} for reviews). In
the majority of investigations the NJl model have been considered
in the mean-field approximation. The successes in phenomenological
applications have stimulated an analysis of a structure of the NJL
model beyond the  mean-field approximation, i.e., in the
next-to-leading orders of mean-field expansion. Such analysis is
necessary for a clarification of the region of applicability for
results and its stability with regard to a variation of parameters
and quantum fluctuations due to higher-order effects.

In this talk I report some results of  the investigation of the
NJL model with the dimensional (analytical) regularization  in the
next-to-leading order of the mean-field expansion.

Due to non-renormalizability a regularization is an  essential
aspect of the NJL model. In contrast to renormalized models, a
parameter of regularization in the NJL model enters the physical
quantities and it is one of essential parameters of the model. For
this reason the least common regularization for NJL model is a
dimensional regularization, since
 the parameter of dimensional regularization traditionally is
treated as a deviation on the  physical dimension of a space and
does not permit any physical interpretation in this treatment.
 However, an
alternative treatment of the dimensional regularization exists --
as a variant of an analytical regularization. In this treatment
all calculations are made in the four-dimensional Euclidean
momentum space, and the regularization parameter is treated as a
power of a weight function, which regularize divergent integrals.
Such treatment of the  dimensional regularization have been
developed and applied to the NJL model in the mean-field
approximation by Krewald and Nakayama \cite{Krew}.

 A possible treatment of the parameter of this analytical
regularization  is some power of a gluon influence on the
effective four-fermion quark self-action of the NJL model.

 Section 1 is based on results of R. Jafarov and myself \cite{JaRo}
and is devoted to  calculations of  meson contributions to the
quark chiral condensate of the NJL model.  The pion contribution
is significant, and at a bound of admissible values of the model
parameters it tends to infinity, i.e., the model is unstable with
respect to quantum fluctuations near this bound. For physical
values of the parameters this contribution is $10\div 20$\% of the
 leading contribution. The sigma-meson contribution is
small for admissible values of the parameter.

In Section 2 non-pole contributions to the chiral condensate are
estimated. These contributions are found to be small. Also in this
section the interesting phenomenon of a cancellation in the scalar
amplitude is discussed: at some value of the analytic
regularization parameter almost all contributions to the scalar
amplitude are cancelled. This effect leads to a "disappearance" of
the sigma-meson from the spectrum of the analytically regularized
NJL model.

  Quite briefly our result can
be formulated as a following statement: the analytically
regularized NJL model does not contain for physical values of
parameters any pathological quantum fluctuations, connected with
the scalar-meson contributions, though the pion contribution is
significant and should be taken into account  at phenomenological
treatments of NJL-type models.

\section{ Analytically regularized NJL model. A mean-field
expansion and
scalar meson contributions to the chiral condensate}

We consider SU(2) NJL model, which  is a theory of a
 spinor field $\psi(x)$ with two flavors (isospin), $n_c$
colors, and $SU_V(2)\times SU_A(2)$-invariant self-action
$$
{\cal L}_{int}=\frac{g}{2} \biggl[(\bar\psi\psi)^2+(\bar\psi
i\gamma_5 \tau^a\psi)^2\biggr]. \label{LSU2}
$$
Here $\tau^a$ are Pauli matrices.

Quark propagator $S^{(0)}$ in the leading mean-field approximation
(Hartree approxomation) is
$$
S^{(0)}= \frac{1}{m-\hat p},
$$
where the  dynamical mass $m$ is a solution  of gap equation
\begin{equation}
1= -8ign_c\int\frac{d\tilde p}{m^2-p^2}. \label{gap}
\end{equation}
Here and below $d\tilde p \equiv d^4p/(2\pi)^4$.
 The divergent
integral in r.h.s. should be considered as a some regularization.

The leading approximation chiral condensate  is
\begin{equation}
\chi^{(0)}=i\tr S^{(0)}(0)=-\frac{m}{g}. \label{chi0}
\end{equation}

The  two-particle amplitude in the mean-field approximation is

\begin{equation}
A = \textsc{1}\otimes\textsc{1}A_\sigma + \tau^a\otimes\tau^a
A_\pi  \label{A} \end{equation}

Here  scalar amplitude $A_\sigma$ and  pseudoscalar amplitude
$A_\pi$ are
$$
A_\sigma=-\frac{ig}{1-L_S},\;\;\;A_\pi=\frac{ig}{1+L_P}
$$
where $ L_S$ and $L_P$ are  scalar and pseudoscalar fermion loops.

Taking into account gap equation (\ref{gap})  one can obtain  the
following representations for $A_\sigma$ and $A_\pi$ in momentum
space:
\begin{equation}
A_\sigma(p)=\frac{1}{4n_c(4m^2-p^2)I_0(p^2)} \label{A_sigma}
\end{equation}
\begin{equation}
 A_\pi(p)=\frac{1}{4n_cp^2I_0(p^2)}.
\label{A_pi}
\end{equation}
Here
\begin{equation}
 I_0(p^2)=\int d\tilde
q\frac{1}{(m^2-(p+q)^2)(m^2-q^2)}. \label{I0}
\end{equation}

Meson contributions to the chiral condensate can be  calculated in
the next-to-leading term of the mean-field expansion.  A
systematic construction of the mean-field expansion can be made by
using the bilocal source formalism. We shall follow for this
purpose to the method of an iterative solution of the
Schwinger-Dyson equation for the generating functional of Green
functions (see \cite{Ro1} for a brief review of the method and
\cite{JaRo} for details of the mean-field expansion in the NJL
model).

The next-to-leading correction to  propagator  $S^{(1)}$ in
mean-field expansion  are defined by the equation for first-step
mass operator $\Sigma^{(1)}=[S^{(0)}]^{-1}\star S^{(1)}\star
[S^{(0)}]^{-1}$ (in x-space):
\begin{equation}
\Sigma^{(1)}(x) =ig\delta(x)\tr S^{(1)}(0)+
S^{(0)}(x)A_\sigma(x)+3S^{(0)}(-x)A_\pi(x). \label{Sigma}
\end{equation}

Due to the non-renormalizability the NJL model should be
regularized, and the regularization is an essence component of the
model. We shall use some special variant of the dimensional
regularization, which was proposed in work \cite{Krew}. In this
approach the dimensional regularization is, in essence, a variant
of the analytical regularization.

Let us consider the approach for the gap equation of the NJL model
as an example. The  gap equation in the Euclidean space after an
angle integration is
$$
1=4gn_c\frac{\Omega_4}{(2\pi)^4}\int \frac{q^2_edq^2}{m^2+q^2_e}.
$$

In correspondence with the prescriptions of \cite{Krew} we modify
the integrand by the weight function
$$
 w_D(q^2_e)
=\Biggr(\frac{\mu^2}{q^2_e}\Biggl)^{2-D/2}
$$
and rescale parameter $\mu^2$ as
$$
(\mu^2)^{2-D/2}=\frac{\Omega_D}{\Omega_4}\frac{(2\pi)^4}{(2\pi)^D}
(M)^{2-D/2}.
$$
where $ \Omega_D=2\pi^{D/2}/\Gamma(D/2). $ Then we obtain for the
gap equation
$$
1=\kappa\Biggr(\frac{m^2}{4\pi M^2}\Biggl)^{D/2-2} \Gamma(1-D/2),
$$
where dimensionless constant $\kappa$ is introduced:
$\kappa=gn_cm^2/2\pi^2.$

 This equation exactly corresponds
to the calculation of the initial integral with the formal
prescription of D-dimensional integration
$$
d\tilde q \equiv \frac{d^4q}{(2\pi)^4}\rightarrow
\frac{(M^2)^{2-D/2}d^Dq}{(2\pi)^D},
$$
but in our case  the calculation was performed in the usual
4-dimensional space, i.e. in our treatment parameter $D$ is not a
dimension of some space, but a parameter which  provides the
convergence. In particular, we do not constrained with the limit
$D\rightarrow 4$ for a treatment of results.

Below we shall use {\it regularization parameter} $\xi$ as $
\xi=(2-D)/2$ and the gap equation has the form
\begin{equation}
1=\kappa\Gamma(\xi)\Biggr(\frac{4\pi M^2} {m^2}\Biggl)^{1+\xi} .
\label{gapxi} \end{equation}
 The region of  convergence of the
integral is $ 0<\xi<1.$ As we shall see, there is also a region
for the physical values of the model parameters.

Integral $I_0$  (see (\ref{I0})), which is a part of scalar
amplitudes $A_\sigma$ and  $A_\pi$, also can be calculated on
above prescriptions. Going to the Euclidean metric, introducing a
standard Feynman parameterization, and translating an integration
variable (which is possible due to  translational invariance of
the procedure, see \cite{Krew}) we can made the angle integration.
In correspondence with the rules, then we introduce  weight
function $w_D(q_e^2)$  and, after the same rescaling, obtain the
result, which also exactly corresponds to the result of an
integration with the formal transition to $D$-dimensional space:
\begin{equation}
I_0(p^2)=\int d\tilde q\frac{1}{(m^2-(p+q)^2)(m^2-q^2)}=
\frac{i\xi \Gamma(\xi)}{(4\pi)^2}\int_0^1 du\Biggl(\frac{4\pi
M^2}{m^2-u(1-u)p^2}\Biggr)^{1+\xi}. \label{I0dim}
\end{equation}

Taking into account gap equation (\ref{gapxi}) we can exclude
$\Gamma(\xi)(4\pi M^2)^{1+\xi}$ from (\ref{I0dim}) and obtain for
$I_0$:
\begin{equation}
I_0(p^2)=\frac{i}{(4\pi)^2} \frac{\xi}{\kappa}\int_0^1
du\Biggl(1-u(1-u)\frac{p^2}{m^2}\Biggr)^{-1-\xi}=
\frac{i}{(4\pi)^2} \frac{\xi}{\kappa}F(1+\xi, 1; 3/2;
\frac{p^2}{4m^2}). \label{I0xi}
\end{equation}
Here $F(a, b; c; z)$ is the Gauss hypergeometric function.

 The  pole terms for the scalar amplitudes in the analytical
 regularization are:
\begin{equation}
A_\sigma^{pole}=
\frac{1}{4n_c(4m^2-p^2)I_0(4m^2)}=\frac{igm^2(1+2\xi)}
{(4m^2-p^2)n_c\xi}, \label{Aspole}
\end{equation}
\begin{equation}
A_\pi^{pole}=\frac{1}{4n_cp^2I_0(0)}= -\frac{igm^2}{p^2n_c\xi}.
\label{Appole}
\end{equation}

 For  pion decay constant
$f_\pi=93$ MeV we obtain very simple formula
\begin{equation}
f^2_\pi=\frac{\xi}{2g}. \label{fxi} \label{fpi}
\end{equation}

We shall use also an expression for  width
$\Gamma_{\pi^0\gamma\gamma}=7.7$ KeV of decay $\pi^0\rightarrow
2\gamma$ (see \cite{Krew}). In our notation this formula is
\begin{equation}
 \Gamma_{\pi^0\gamma\gamma}=\frac{\alpha^2m^3_\pi\xi^2(1+\xi)^2}
 {64\pi^3f^2_\pi\kappa^2}.
\label{Gamma}
\end{equation}

Using these formulae, one can fix  parameters $m, g($or $\kappa)$
and $\xi$ of the analytically-regularized
 NJL model in the leading mean-field approximation.

A new thing in our approach in comparison with that of Krewald and
Nakayama \cite{Krew} is a systematical exploiting of the gap
equation to exclude the dependence of the model parameters on the
dimensional parameter $M$, and,  correspondingly, a significant
simplification of formulae for observable values.

As a measure of quantum fluctuations of the chiral field,
consider a ratio of first-step condensate \\ $ \chi^{(1)}=i \tr
S^{(1)}(0)$
 to
leading-approximation condensate $\chi^{(0)}$:
\begin{equation}
r\equiv \frac{\chi^{(1)}}{\chi^{(0)}}=r_\sigma+r_\pi \label{r}
\end{equation}
Here $r_\sigma$ is a scalar contribution and $r_\pi$ is a
pseudoscalar contribution. Integrals for $r_\pi$ are calculated in
the analytical regularization in closed form and give us a very
simple expression (at $n_c=3$) :
\begin{equation}
r_\pi=\frac{1}{8\xi}. \label{rpi}
\end{equation}
 Scalar contribution $r_\sigma$
also is a function of parameter $\xi$ only and can be represented
as an integral with the Gauss hypergeometric function (see
\cite{JaRo}).
 Our results indicate, that quantum fluctuations of the  chiral
condensate in the  NJL model can be significant at some values of
the regularization parameter $\xi$. The fluctuations caused by
pseudoscalar field are large at $\xi\rightarrow 0$. The
sigma-meson contribution is small in the region of physical values
of the model parameters.

 Numerical results are following:

 At $c=-160$ MeV we have $\xi\simeq 1,\;m\simeq 475$
MeV, $\kappa\simeq 2$. At this value of  $\xi$ the correction to
condensate  is about  3\%,  i.e. the model is stable with respect
to meson fluctuations. But such a low value of the chiral
condensate hardly corresponds to the phenomenology -- it lead to
large current quark masses.

At $c=-200$ MeV we obtain  $\xi\simeq 0.44,\;m\simeq 400$ MeV,
$\kappa\simeq 0.62$. The correction to the condensate   is
 9\%.

 At $c=-250$ MeV we obtain $\xi\simeq 0.2,\;m\simeq 370$ MeV,
 $\kappa\simeq
 0.24$,
 and the correction to the condensate   is more then 18\%.

Here condensate $c=(\chi/2)^{1/3}$.

 The calculated corrections permit us to modify the choice of
 parameters by the following modification of the condensate formula
\begin{equation}
\chi=-\frac{m^*}{g^*}[1+r(\xi^*)].
 \label{chi*}
\end{equation}

This modified choice of parameters give us

-- at $c=-200$ MeV: $\xi^*\simeq 0.56,\; m^*\simeq 420$ MeV,
$\kappa^*\simeq 0.86$; the condensate correction is 7\%;

-- at $c=-250$ MeV: $\xi^*\simeq 0.3,\; m^*\simeq 380$ MeV,
$\kappa^*\simeq 0.39$; the condensate correction is 13\%;

Apparently, the modification decreases the fluctuation of
condensate, i.e. it stabilizes the situation.

\section{Beyond the pole approximation. Cancellations in the
scalar amplitude}

The calculations of Section 1 were performed with pole
approximations (11) and (12) for amplitudes $A_\sigma$ and
$A_\pi$. Pole approximation (12) for pseudoscalar amplitude
$A_\pi$ is well-defined for the admissible values of
regularization parameter $\xi$. But similar approximation (11) for
scalar amplitude $A_\sigma$ is a questionable thing.

The region of the convergence for integral $I_0$ is $-1<\xi<1$.
Hypergeometric function $F(1+\xi, 1; 3/2; z)$  (see eq.
(\ref{I0xi})) is singular in point $z=1$ (i.e., at $p^2=4m^2$) at
$\xi>-1/2$. This threshold singularity is an artefact of the
regularization, and, consequently, we should consider pole
approximation (12) of the scalar amplitude as an analytic
continuation from region $-1<\xi<-1/2$ to physical region
$0<\xi<1$. This analytic continuation seems to be necessary for a
physical interpretation of the sigma-meson as a particle in the
spectrum of the NJL model.

From other side, if one  considers scalar amplitudes $A_\sigma$
and $A_\pi$ beyond the pole approximation, then an  interesting
phenomenon in the analytically regularized NJL model arises. At
$\xi=1/2$ the hypergeometric function in  eq.(\ref{I0xi}) reduces
to a simple elementary function: $F(3/2, 1; 3/2; z)=(1-z)^{-1}$,
and we obtain:
\begin{equation}
I_0\mid_{\xi=1/2}=\frac{i}{4gn_c(4m^2-p^2)},
\end{equation}
and, in correspondence with eqs. (\ref{A_sigma}) and (\ref{A_pi}):
\begin{equation}
A_\sigma=-ig,\;\; A_\pi=ig-\frac{4igm^2}{p^2}. \label{A1/2}
\end{equation}
These exclusively simple formulae for the amplitudes demonstrate
an almost full cancellation of contributions to scalar amplitude
$A_\sigma$ at $\xi=1/2$
 ( except of the leading
order of perturbation theory). One of the consequences of such
cancellation is a "disappearance" of the sigma-meson from the
spectrum of the NJL model.

Formulae (\ref{A1/2}) for the amplitudes enable also to estimate
the corrections to the chiral condensate beyond the pole
approximation. At $\xi=1/2$ by using amplitudes (\ref{A1/2}) we
obtain
\begin{equation}
r_\sigma=0,\;\;r_\pi=\frac{1}{4}.
\end{equation}
Apparently the pseudoscalar contribution is exactly the same as
the pole contribution of eq.(\ref{rpi}). Hence, non-pole
contributions at $\xi=1/2$ disappear:
$r_\sigma^{non-pole}=r_\pi^{non-pole}=0.$
This exact result gives us a possibility to estimate non-pole
contributions near this value of regularization parameter $\xi$
(i.e., in whole region of the physical values of the model
parameters) as a little.\\

I thank the Organizing Committee  to make my participation in
QFTHEP'2003 possible.

\end{document}